\documentclass[twocolumn, twoside]{IEEEtran} %!PN

\usepackage{epsfig}
\def\e{\begin{equation}}
\def\f{\end{equation}}
\def\=#1{\overline{\overline{#1}}}
\def\-#1{{\bf #1}}
\def\oo{\omega}

\def\.{\cdot}
\def\x{\times}

\def\r#1{(\ref{eq:#1})}

\def\a{\end{array}\right)}
\def\lab#1{\label{eq:#1}}
\def\r#1{(\ref{eq:#1})}

\begin{document}
\title {Backward wave region and negative material parameters of a structure formed
by lattices of wires and split-ring resonators }
\author{Constantin R. Simovski, Pavel A. Belov, {\it Student\ Member IEEE}, and Sailing He
\thanks{C. Simovski is with the Physics Department, St. Petersburg Institute of Fine Mechanics and Optics, Sablinskaja Street, 14, St. Petersburg, 197101, St. Petersburg, Russia. E-mail: simovsky@phd.ifmo.ru.}
\thanks{P. Belov is with the Radio Laboratory, Helsinki University of Technology, P.O. Box 3000, FIN-02015, Espoo, Finland. E-mail: belov@rain.ifmo.ru.}
\thanks{Sailing He is with Centre for Optical and Electromagnetic Research, State Key Laboratory of Modern Optical Instrumentation, Zhejiang University, Yu-Quan, 310027 Hangzhou, P. R. China and with Division of Electromagnetic Theory, Alfven Laboratory, Royal Institute of Technology,  S-100 44, Stockholm, Sweden.}}
\markboth{IEEE Transactions on Antennas and Propagation, Specila Issue on Metamaterials, 2003}
{Simovski, Belov and He: Backward wave region and negative material parameters } %!PN

\maketitle

\begin{abstract} A structure formed by combined lattices of infinitely
long wires and split-ring resonators is studied.
A dispersion equation is derived and then used to calculate
the effective permittivity and permeability in the
frequency band where the lattice can be homogenized.  The backward
wave region in which both the effective permittivity and
permeability are negative is analyzed. Some open and controversial
questions are discussed. It is shown that previous experimental
results confirming the existence of backward waves in such a
structure can be in deed explained in terms of negative material
parameters. However, these parameters are not quasi-static and
thus the known analytical formulas for the effective material
parameters of this structure, which have been widely used and
discussed in the literature, were not correct, and it was the
reason of some objections to the authors of that experiment. 
\end{abstract}

% ************* Begin of the body of the manuscript **********

\section{Introduction}

Meta-materials with negative permittivity and 
negative permeability,which were first suggested in \cite{Veselago}, have attracted much
attention recently. A meta-material which has
simultaneously negative permittivity $\epsilon$ and 
negative permeability $\mu$ within a certain
frequency band at microwave frequencies  has been 
introduced recently \cite{negative}. This structure
consists of two lattices: a lattice of
infinitely long parallel wires, and a lattice of relatively
small (compared to the wavelength $\lambda$ in the host
medium) particles  which are called {\it split-ring resonators} (SRR:s). 
In \cite{Kostin} and \cite{Pendry} two analytical
models of SRR (similar to each other) have been developed for the resonant permeability at
microwave frequencies. Lattices of wires at low frequencies (when the lattice
period $d$ is smaller than $\lambda/2$) were considered as
homogenuous dielectric media  long time ago in \cite{Brown} and 
were studied again recently in the low-frequency region
\cite{Pendry2,StasMOTL}. At these low frequencies the negative permittivity
is due to the lattice of wires according to the
models of \cite{Brown,Pendry2,StasMOTL} (for waves propagating
normally to the wires with the electric field polarized along
these wires). These results were combined in \cite{negative,Smith,smithclc,Shelbyap} 
to form a simple model predicting simultaneously
negative $\epsilon$ and $\mu$ within the resonant band of a SRR.
In \cite{Smith} this prediction
has been qualitatively confirmed by numerical simulations using the
MAFIA code. Dispersion curves obtained numerically contain the
pass-band within the SRR resonant band (due to the
presence of the SRR lattice). This pass-band can also be predicted by the analytical model. However, the numerical dispersion data obtained in \cite{Smith} have not been  used to extract the material
parameters.
The
experimental observation of the negative refraction of a wave in such
a structure is reported in  \cite{Science}. The phenomenon of the negative refraction was predicted
in \cite{Veselago} for media with $\epsilon<0$ and $\mu<0$, and according to this theory  they correspond to
 the backward wave
region  (where the Poynting vector of the eigenwave is
opposite to the wave vector).

Does the experimental observation of \cite{Science} mean that the structure
suggested in \cite{negative} can be described through $\epsilon$
and $\mu$ which are both negative within the SRR resonant band?
Based on \cite{Science} one can only assert that the backward
wave region necessarily exists within this frequency band. Backward waves in a lattice
correspond to negative dispersion, i.e., the group velocity
(the derivative of the eigenfrequency with respect to the wave number) is in the opposite 
direction of the phase velocity. Negative
dispersion for a lattice is quite common in high frequency bands
($kd>\pi$, where $k=2\pi/\lambda$ is the wave number in the host
medium). However, at low frequencies ($kd<\pi$) negative dispersion is
an abnormal phenomenon. When $kd<\pi$ the wavelength
$\lambda$ is high enough as compared to the lattice period and the SRR
size and thus the homogenization of the lattice may be
possible. In \cite{negative},\cite{Science} and
\cite{Smith} the lattice
at low frequencies was treated as a continuous medium and the concept of the negative
dispersion is equivalent to the concept of negative material
parameters. However, is it possible to describe the structure formed by lattices of 
wires and SRR:s (studied in \cite{negative} and \cite{Science})
in terms of  $\epsilon$ and $\mu$ within the resonant band of SRR:s? This
question remains open since the analytical model \cite{negative}
with which these material parameters were introduced is incomplete
and can be wrong.

In \cite{Science} the negative refraction was observed only for
waves whose  electric field is parallel to the wires and the wave vector is 
 perpendicular to the wires. Thus, the permittivity
$\epsilon$ considered in \cite{Pendry2} and \cite{negative} are the
$xx-$component of the permittivity tensor (we assume  the wires are
along the $x$-axis) and the permeability $\mu$ considered there are  the transversal component of the permeability
tensor $\mu_t=\mu_{yy}=\mu_{zz}$ (there are two 
orthogonal SRR:s in each unit cell of the structure studied in
\cite{Science}). If one considers only the propagation
in the transversal plane (i.e., the $y-z$ plane in our case), one can neglect
the spatial dispersion in the wire lattice and consider it as a
medium with negative permittivity \cite{Brown}. In the case of perfectly conducting wires one has \cite{Brown}: \e
\epsilon=1-{\oo_p^2\over \oo^2} \lab{brown}\f where $\oo_p$ is the
analogue of the plasma frequency. In \cite{Pendry,Pendry2,Shelbyap}
the following frequency dependencies of the effective medium parameters  were suggested for their structure when both the SRR particles and the wires are made of real
metal: \e \epsilon=1-{\oo_p^2-\oo_e^2\over \oo^2-\oo_e^2-j\gamma \oo}
\lab{Pend1}\f \e \mu=1-{\oo_{mp}^2-\oo_m^2\over \oo^2-\oo_m^2-j\gamma
\oo} \lab{Pend2}\f where $\oo_{mp}$ is the analogue of the resonant frequency
of a magnetic plasma, $\oo_{e}$ is the electronic resonant
frequency and $\oo_m$ is the magnetic resonant frequency
(i.e.,, the resonant frequency of the magnetic
polarizaibility for an individual SRR). These formulas correspond
to the conventional quasi-static model for the permittivity and
permeability, which takes into account only the local interaction
in the structure under the homogenization (i.e., the electromagnetic interaction between the SSR:s and wires are neglected).  In the present paper, we take into account the electromagnetic interaction (including the plane-wave interaction)  between the SSR:s and the wires. We consider the
lossless case (corresponding to $\gamma=0,\ \oo_e=0$ in \r{Pend1} and
\r{Pend2}).

In the present paper, we also discuss  the following two questions
:
\begin{itemize}
\item Is it possible to consider the structure formed by lattices of wires and SRR:s
as a medium that its $\epsilon$ is the same as the effective
permittivity of a wire medium and its $\mu$ is the same as the effective
permeability of the lattice of SRR:s (i.e., is it possible to neglect
the electromagnetic interaction between the SRR:s and the wires when calculating the
material parameters of the whole structure)?

\item How to find the frequency region 
within the resonant band of SRR scatterers
in which the homogenization of the whole structure is allowed? 
At some frequencies within the SRR resonant band the homogenization model may give  
 so high values of the effective permeability 
that the product $|\epsilon\mu|$ becomes very large. Though the wavelength $\lambda$
in the host medium is much larger than the lattice
period $d$, the effective wavelength $\lambda/\sqrt{|\epsilon\mu|}$ of the eigenmode at these frequencies
 can be of
the order of $d$ or even smaller  and thus the homogenization model becomes
contradictory. However, within the 
band of the resonance there are frequencies at which the product 
$|\epsilon\mu|$ is not so large. Thus , the problem is how to separate quantitatively the  
frequencies at which the homogenization is forbidden from the frequencies at which 
the homogenization  is allowed.

\end{itemize}

The first question has been considered briefly in \cite{book}. It was
indicated that the structure considered in \cite{Science} was fortunately
built so that each SRR is located exactly at the center of
two adjacent wires and thus there is no quasi-static interaction
between the wire lattice and the SRR lattice (i.e., the magnetic fields produced by
the two adjacent equivalent line currents cancel out at the center where  the SRR is
located). If one considers
$\epsilon$ and $\mu$ as quasi-static parameters (as was done in \cite{book}), the absence of the quasi-static
interaction should lead to the following result: the effective permittivity of the structure  is identical to the effective permittivity of the lattice of wires and the effective permeability of the structure  is identical to the effective permeability  of the lattice of SRR:s.
However, we will show in the present paper that this is not correct
since the electromagnetic interaction between the wires and the SRR:s in such a structure is not quasi-static (or local) and will 
dramatically influence the  effective permittivity.
The consideration of structure as lattice of wires 
positioned in a host negative magnetic suggested in \cite{Utah} 
is not adequate, because it also does not describe electromagnetic interaction correctly.

For the second question, one must be
very careful in the homogenization of the complex structure studied in \cite{negative}, \cite{Science} and \cite{Smith}.
 In fact, the 
 results of \cite{Science} can not be interpreted quantitatively in terms of
the permittivity and permeability used in \cite{Pendry} and
\cite{Pendry2}, \cite{negative}, \cite{Science}
and \cite{Smith}. This has been revealed in \cite{Garcia}.

In the present work we develop an analytical model for  a
structure similar to the one  studied in \cite{Science} (i.e., formed by combined lattices of infinitely long wires and split-ring resonators). The model allows
the structure to be homogenized and its valid frequency domain to be identified. A self-consistent dispersion equation is derived 
and then used to calculate correctly the effective permittivity
and permeability in the frequency band where the lattice can be
homogenized. The backward wave region   (which is a part of the
resonant band of SRR) in which both the effective
permittivity and permeability are negative is analysized.
Since our homogenization  is based on 
the dispersion characteristics of the structure, the extracted 
material parameters are not quasi-static (i.e., not
only the local but also the non-local interactions are considered among the elements of the
structure). Our results have shown  that the homogenization is allowed
only over part of the resonant band of the SRR scatterers and the homogenization is forbidden in 
a sub-band inside the SRR resonant band (even though the frequencies of this sub-band
are low and the spatial dispersion exists there).

\section{SRR with identical rings}

The SRR particle considered in \cite{Kostin} and \cite{Pendry} is a pair of
two coplanar broken rings. Since the two loops are not identical the
analytical model for this particle  is rather
cumbersome (SRR models
more complete than those suggested in \cite{Kostin} and \cite{Pendry}
have been developed in \cite{Bruno} and \cite{Marques}). It is not correct that the SRR particle can be
described simply as a resonant magnetic scatterer \cite{Marques}.
The structure considered in \cite{Science} (if homogenized) has to be
described through three material parameters: $\epsilon$, $\mu$
and $\kappa$ (the magnetoelectric coupling parameter). It was
shown in \cite{Marques} that a SRR is actually a bianisotropic
particle and the role of bianisotropy is
destructive for negative refraction.

The bianisotropy is not the only disadvantage of this SRR
particle. Another disadvantage is its resonant electric
polarizability (i.e., the polarization produced by the electric field), which was
also ignored in \cite{Kostin} and \cite{Pendry}. Electric
polarizability resonance occurs at frequencies very close to the resonant frequency of the
magnetic polarizability \cite{Bruno}. If the SRR:s are made of real
metal, the resonant electric
polarizability  will lead to a dramatic increase in the resistive loss. The resonant electric
polarizability  also
makes the analytical modelling of the whole structure very
complicated.

\begin{figure}
\centering \epsfig{file=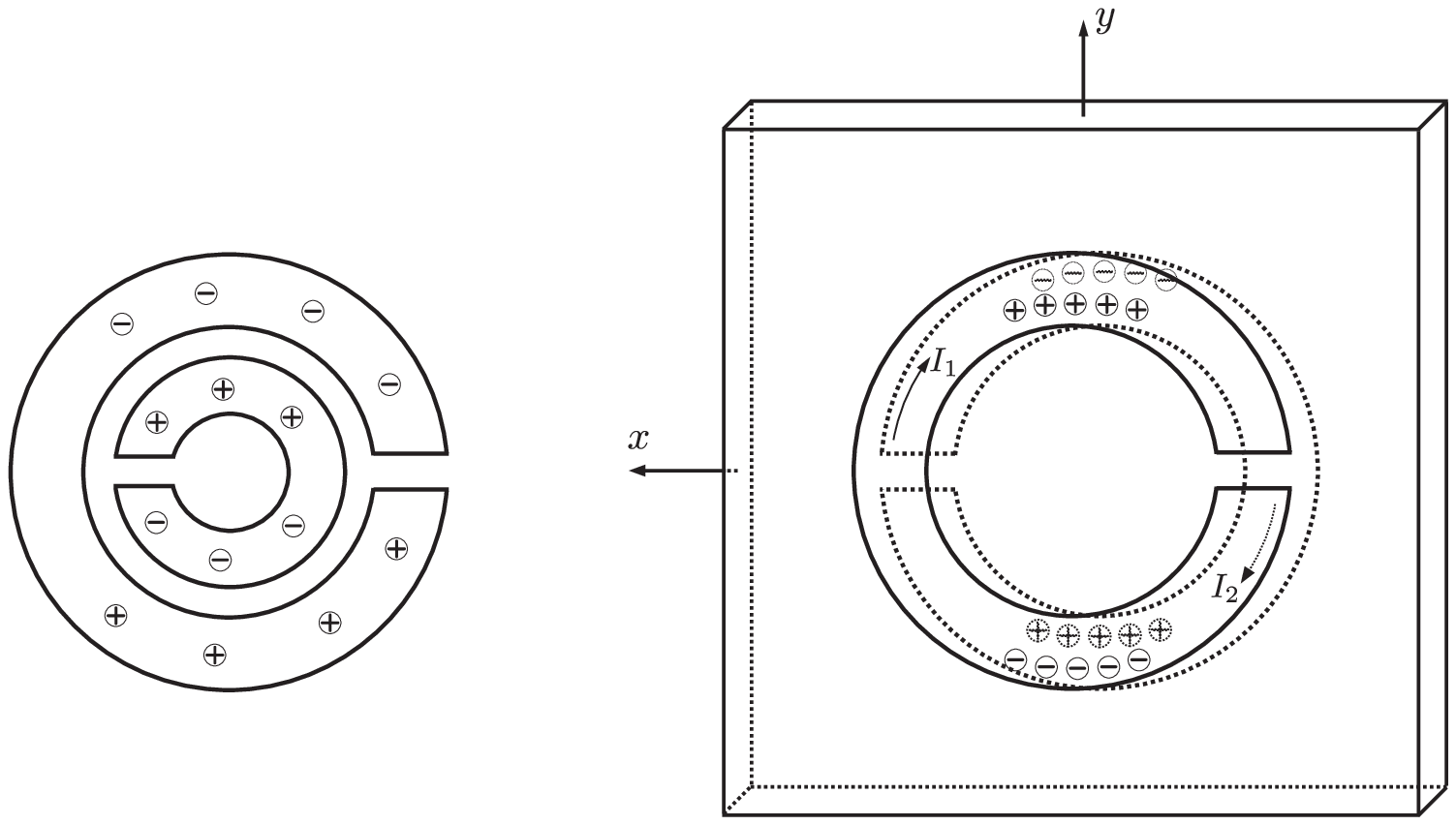, width=9cm} 
\caption{Left: coplanar SRR.
%from considered in \cite{Kostin} and \cite{Pendry}.
Right: parallel SRR.
%from introduced in \cite{Marques}. Concept 
The concept of the mutual capacitance is illustrated by indicating the
charges in the polarized rings.} 
\label{marques}
\end{figure}

 A  modified  SRR which does not
possess bianisotropy  was proposed in \cite{Marques}.  This SRR also consists of two
loops but they are identical and parallel to  each other (located on both sides of a dielectric plate in
practice). Fig. \ref{marques} shows two kinds
of SRR. The left one  is the
SRR considered in  \cite{Kostin} and \cite{Pendry}. The right is
the SRR introduced in \cite{Marques}  and the one considered in the present paper. It has been mentioned  in
\cite{Marques} that the magnetic resonant frequency of their SRR is lower than that of  the SRR considered in  \cite{Kostin} and \cite{Pendry}
 (for the same size). 
This is because  
the mutual capacitance
$C_{\rm mut}$ between the two parallel broken
rings is now the capacitance of a conventional parallel-plate
capacitor and is significantly higher than the mutual
capacitance of two coplanar split rings considered in \cite{Pendry}. This fact
is illustrated in Fig.~\ref{marques}: if the upper half of ring 1
is charged positively the negative induced charges appear in the
upper half of ring 2. The same situation happens for the SRR shown in the left
part of this figure, but not so effectively since the strips are
coplanar and weakly interacted. Therefore, the homogenization is more appropriate 
within the resonant band of the particle since the ratio of the particle size to $\lambda$
is small.

There is one more advantage that was not mentioned in
\cite{Marques} for the SRR of two parallel broken
rings. The resonances of the
electric and magnetic polarizabilities of this particle do not
overlap in frequency. 
Actually, particle suggested in \cite{Marques} is  a special case
of the bi-helix particle introduced (with the
aim to create novel low-reflective shields) and studied in
\cite{electromagn}. Unlike the SRR considered in 
\cite{Marques} this bi-helix particle contains four stems orthogonal to the loop planes. Note that
the theory of \cite{electromagn}
remains valid even if the length of these stems becomes zero.

Both rings 1 and 2 (see the right diagram of Fig. \ref{marques}) have the
same radius $r$ and area $S=\pi r^2$. 
The impedance $Z_0$ for each broken ring can be calculated by
$$
Z_0={1\over Y_{\rm ring}}+{1\over Y_{\rm split}},
$$
where $Y_{\rm ring}$ is the admittance for the corresponding closed ring 
and $Y_{\rm split}$ is the admittance for the split (associated with  the
capacitance between the two broken ends). The magnetization arises due to the magnetic field
orthogonal to the ring plane (i.e., the $xy$ plane in Fig. \ref{marques}).
Resonant electric polarization is caused by the $y-$component of
the external local electric field $\-E^{\rm loc}$. The $x$-component of $\-E^{\rm loc}$ has no influence over the resonant polarization and can be neglected
\cite{IEEE}. Thus, voltages
(electromotive forces) ${\cal  E}^{\rm H}$ and ${\cal  E}^{\rm E}$  will be 
induced in each loop by the external local electric and
magnetic fields, respectively  \cite{electromagn}, \cite{IEEE}: 
\e {\cal  E}^{\rm
H}=-j\oo \mu_0 SH_z^{\rm loc} \lab{mag}, \f
 and
 \e
{\cal  E}^{\rm E}={4r J'_1(kr)Z_0\over j\eta A_1(kr)}\left( 1+
{j\over \pi\eta(Y_{\rm ring}+Y_{\rm split})} \right)E_y^{\rm loc} \lab{el},\f
where $J'_1$ is the derivative (with respect to the argument
) of the Bessel function, and $A_1$ is one of the so-called King's coefficients known in the theory of loop antennas (see e.g. \cite{IEEE}). 
Then the induced currents $I_{1,2}$ (due to the changing of the charges at the tips of the  split arms) at the split gaps of
rings 1 and 2 satisfy the following equations,
\cite{electromagn}:
\begin{eqnarray}
I_1Z_0+I_2Z_{\rm mut} &=&{\cal  E}^{\rm E}   +{\cal  E}^{\rm H},
\lab{ppp} \\
I_2Z_0+I_1Z_{\rm mut}  & =  &{\cal  E}^{\rm E}   +{\cal  E}^{\rm H},
\lab{mmm}
\end{eqnarray}
where $Z_{\rm mut}$ is the mutual impedance of the  two broken  rings.

It is clear from \r{ppp} and \r{mmm} that there are
two eigenmodes of currents in the SRR. The first mode corresponds to
$I_1=I_2$ when the electric dipole moment of the SRR is zero and the magnetic dipole moment $\-m=m\-z_0$ 
 is twice the magnetic moment of a single 
ring (see Fig. \ref{marques}). The second mode
corresponds to $I_1=-I_2$  when the magnetic moment is zero and the electric dipole moment
$\-p=p\-y_0$ of the SRR is twice the electric dipole moment of
a single ring. The first mode in the SRR is excited by the local magnetic field, and 
the
second mode is exceited by the local electric field. The more the two rings are mutually coupled,  the more is
the difference between the resonant frequencies of the electric and
magnetic  resonantors \cite{electromagn}. The electric resonant frequency (at which the electric
polarizability resonates)  is always higher
than the magnetic resonant frequency \cite{electromagn}.
The relative difference of these two resonant frequencies may exceed $50\%$
\cite{electromagn}. If the distance $h$ between the two parallel broken rings 
is very small, we can assume that the mutual coupling is so strong
that the electric polarizability of the SSR is negligible
within the frequency band of the magnetic resonance.  Note that $h$ is also the thickness of
the dielectric layer between the two parallel broken rings in Fig. \ref{marques} since the rings are
assumed to be perfectly conducting with infinitesimal thickness.

Therefore,  unlike the SSR considered in
\cite{Kostin} and \cite{Pendry},  the SSR suggested in \cite{Marques} is appropriate for creating
artificial magnetic resonance   without resonant electric properties
within the frequency band of interest. This is the reason why we choose the SSR shown in the right part of
Fig.~\ref{marques} to study in the 
present paper (the analysis will be much simpler).  

In the present paper we introduce  an analytical model for an individual SRR particle
used in  \cite{Marques}. This model is simpler than  the one considered
in \cite{electromagn} (due to the absence of the stems). Assume that
the SRR shown in Figure \ref{marques} is made of perfectly
conducting strip and is excited by magnetic field $H_z^{\rm loc}$.
Also assume that the dielectric plate separating the two parallel rings has
the same permittivity as the background medium (then we can avoid
the influence of the dielectric plate which can be
very strong if there is a mismatch in the permittivity). The model used in this section  is
quasi-static since it refers to an isolated particle of small size
(with respect to the wavelength).

If the non-uniformity of the azimuthal current distribution
in both rings and SRR:s can be neglected, the magnetic polarizability can be written as  \e
a_{mm}={m\over H_z^{\rm loc}}={2I\mu_0S\over H_z^{\rm loc}}
\lab{def}\f where $I=I_1=I_2$. From \r{ppp} and \r{mag} it follows that 
\e I={{\cal E}^{\rm H}\over Z_0+Z_{\rm mut}}={-j\oo S\mu_0H_z^{\rm
loc}\over Z_0+Z_{\rm mut}} \lab{tok}\f

Calculating the total impedance of the loop by taking into account the
mutual coupling of the loops (as it was done in \cite{Bruno} and \cite{Marques}  for SRR of coplanar
rings), we obtain
$$
Z_{\rm tot}=Z_0+Z_{\rm mut}=j\oo (L+L_{\rm mut})+{1\over j\oo C_{\rm tot}}+R_r
$$
Here $R_r$ is the radiation resistance of the whole particle, $L$
is the ring inductance:
$$
L=\mu_0r\left(\log{32r\over w}-2\right)
$$
where $w$ is the width of the strip from which the ring is
made of, and $L_{\rm mut}$ is the mutual inductance of the two parallel
coaxial rings:
$$
L_{\rm mut}=\mu_0r\left[\left(1+{3\xi^2\over 4}\right)\log{4\over
\xi}-2\right]
$$
where $\xi=h/2r$. The total capacitance $C_{\rm tot}$ attributed to
the split 
can be calculated (taking into account the capacitive mutual coupling;  cf. \cite{Bruno} and
\cite{Marques}) as half of the mutual capacitance formed by the two
parallel rings
$$
C_{\rm tot}={C_{\rm mut}\over 2}={\epsilon_0\epsilon w\pi r\over 2h}
$$
In this formula the capacitance of the split is neglected
since it is small as compared to $C_{\rm mut}$.

From \r{def} and \r{tok} we obtain \e a_{mm}= {2\mu_0^2S^2\over
(L+L_{\rm mut})\left({\oo_0^2\over \oo^2}-1\right)-j{R_r\over\oo}}
\lab{amm}\f where \e \oo_0^2={1\over C_{\rm tot}(L+L_{\rm mut})} \lab{omega}\f
In a similar way one can show that the electric polarizability
resonates at the frequency $\oo_1$ (see also 
\cite{electromagn}):
$$
\oo_1^2={1\over C_{\rm tot}(L-L_{\rm mut})}
$$
and $\oo_0<\oo_1$.

We will also use the following result of  \r{amm}: \e {\rm
Re}\left({1\over a_{mm}}\right)={(L+L_{\rm mut})\left({\oo_0^2\over
\oo^2}-1\right)\over 2\mu_0^2S^2} \lab{inverse}\f

The radiation resistance  $R_r$ can be found from the following condition
\cite{Sipe,nonresPRE,JOSA}: \e {\rm Im}\left({1\over a_{mm}}\right)={k^3\over
6\pi \mu_0} \lab{ima}\f
In the dispersion equation for a lattice, $R_r$ cancels out and does
not influence the result.

\section{The structure }

The structure we study in the present paper is similar to the one  studied experimentally
in \cite{Science}, however, instead of the coplanar SRR:s we use the parallel
SRR:s (as described in the previous section).

\begin{figure}
\centering \epsfig{file=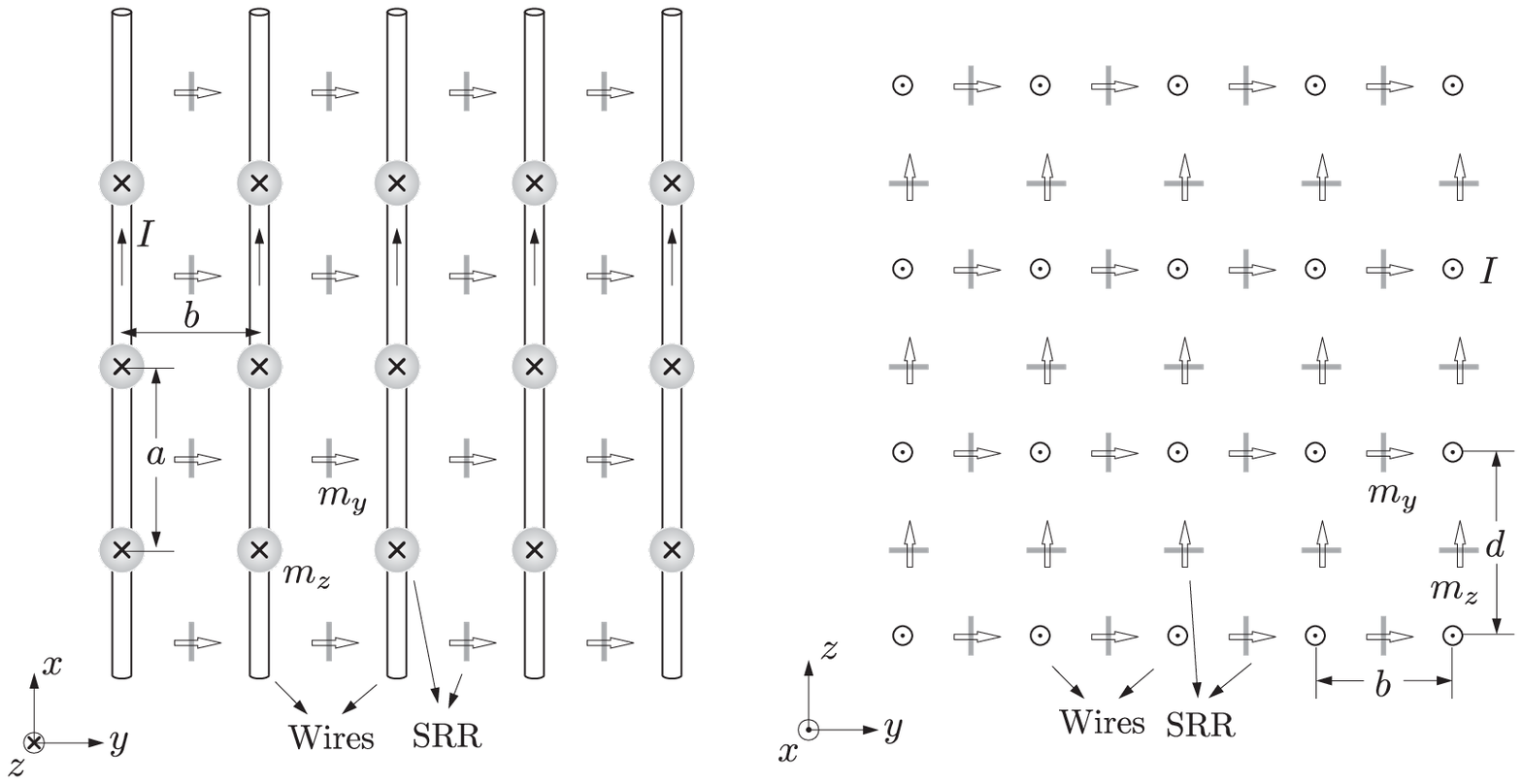, width=9cm} \caption{ Left:
front view of the lattice of SRR:s (shown as disks; their
magnetic dipoles are indicated with arrows) and straight wires.
Right: top view. The wave propagation is in the $y-z$ plane. }
\label{grid}
\end{figure}

When the wave propagates along the $z$ axis the electric field excites
the $x-$directed current $I_{n_y,n_z}$ in the wire  numbered
$(n_y,n_z)$ (for the reference wire we have $n_y=n_z=0$). The magnetic field
excites those SRR:s which are parallel to the $x-z$ plane. Their
magnetic moments are parallel to the $y$ axis. Then the lattice can
be considered as a set of 2D grids parallel to the $x-y$ plane and
orthogonal to the propagation direction. Each grid contains
magnetic and electric polarizations. The magnetic moments as well as
the currents are tangential to the grid plane, and each grid can be
considered as a sheet of surface magnetic moment $\-M=M\-y_0$ and
a surface electric current $\-J=J\-x_0$ (or surface electric
polarization $P=J/j\oo$). Similar situation holds for the wave 
propagation along the $y$ axis. Then one has $\-M=M\-z_0$ and
$\-J=J\-x_0$, and the electric and magnetic polarizations for each
2D grid (parallel to the $x-z$ plane) is again tangential to the grid
and orthogonal to the propagation direction. The wire
lattice and the SRR lattice have the same periods along the $y$ and $z$ axes 
and are denoted as $b$ and $d$, respectively. The period of the SRR lattice
along the $x$ axis is denoted as $a$. Fig.
\ref{marques} shows the two orthogonal sets of SRR:s separated with each other 
by $a/2$ along the $x$ axis. This separation plays no role in our
model  since we do not consider the electromagnetic interaction between
these two orthogonal sets of magnetic dipoles. When the wave
propagates along the $z$ axis or $y$ axis, one of the two sets of SRR:s
is not excited and the interaction is completely absent.

In the case $b=d$ such a structure behaves (within the frequency
band where the homogenization is possible) like a uniaxial
magneto-dielectric medium with relative axial permittivity
$\epsilon_{xx}$ (mainly due to the presence of the wires) and
relative transversal permeability $\mu_{yy}=\mu_{zz}=\mu_{t}$
(mainly due to the SRR particles). This indicates  that in order to find the
effective material parameters of the whole structure we can consider only the
case of the normal propagation (along the $z$ or $y$ axis).

Note that the transversal permittivity and the axial permeability of the structure
are equal to those of the background medium. For
simplicity we assume this background medium is vacuum.

We will see that the SRR:s strongly interact with the wires at each
frequency. Their interaction is not quasi-static  and 
influences the propagation constant  starting from zero
frequency. In this way it influences the material parameters of
the whole structure.

\section{Dispersion equation}

Let the wave propagate along the $z$ axis with propagation factor
$\beta$ (to be determined). Consider the whole structure as a set of
parallel 2D grids which are parallel to the $x-y$ plane and 
denote the surface magnetic moments
$M(n_z)$ and the surface currents $J(n_z)$ at those grids numbered $n_z$. Then we choose  an arbitrary
SRR in the grid with $n_z=0$ as the reference particle and
an arbitrarily chosen wire (in the same grid) as the reference wire.

When we evaluate the magnetic moment $m$ of the reference SRR
(which is related to the surface magnetic moment $M$ by $m=Mab$),
we take into account its electromagnetic interaction with all the 
other SRR:s following the work of \cite{JOSA} where a simple model
of 3D dipole lattice was suggested. As to the influence of the
wires to the reference SRR, we can replace each grid of wires with
a sheet of current $J(n_z)$ because of the absence of the
a quasi-static interaction between SRR:s and wires. This
gives the well-known plane-wave approximation of
the electromagnetic interaction in lattices (see \cite{JOSA}
and the references cited there).

When we evaluate the current $I$ of the reference wire (which is
related to the surface electric current by $I=Jb$), we take into
account its interaction with all the other wires following the work of
\cite{Belov} where a simple model of doubly-periodic wire
lattice was suggested. The influence of the SRR lattice on the
reference wire  is taken into account under the plane-wave
approximation and the reciprocity principle is satisfied.

Each sheet of electric or magnetic polarization produces a
plane wave \cite{Lindell}.  Since $M(n_z)$ and $J(n_z)$ satisfy
\begin{eqnarray}
M(n_z) &=& Me^{-jn_z\beta d}
\lab{mum} \\
J(n_z) &=& Je^{-jn_z\beta d}
\lab{juj}
\end{eqnarray}
we can write the following relations for the $x-$component of the
electric field (produced by all the sheets of magnetic moment $M$ and
acting on the reference wire) and the $y-$component of the
magnetic field (produced by all the sheets of current $J$ and acting on
the reference SRR):
\begin{eqnarray}
E_x^{M} &=& \sum\limits_{n_z=-\infty}^{\infty}{\rm sign}(n_z){j\oo
M\over 2}e^{-jn_z\beta d-jk|n_z|d}
\lab{eee} \\
H_y^{J} &=& \sum\limits_{n_z=-\infty}^{\infty} {\rm
sign}(n_z){J\over 2}e^{-jn_z\beta d-jk|n_z|d} \lab{hhh}
\end{eqnarray}
Both series can be analytically carried out  and we easily obtain
\begin{eqnarray}
E_x^{M} &=& -{\oo M\over 2}{\sin \beta d\over \cos kd -\cos \beta d}
\lab{esum} \\
H_y^{J} &=& {jJ\over 2}{\sin \beta d\over \cos kd -\cos \beta d}
\lab{hsum}
\end{eqnarray}

The local electric field acting on the reference wire is the sum of
$E_x^{M}$ and the contribution of the wires: \e E_x^{\rm
loc}=E_x^{M}+C_wI=E_x^{M}+C_wbJ \lab{eloc}\f where $C_w$ (the
interaction factor of the wire lattice) was determined in
\cite{Belov}: \e C_w={-j\eta\over 2b}\left[{\sin kd\over \cos kd
-\cos \beta d}+{kb\over \pi}\left( \log{kb\over 4\pi} +\gamma
\right)+j{kb\over 2} \right] \lab{aaa}\f Here $\eta$ is the wave
impedance of the host material and $\gamma=0.5772$ is the Euler
constant.

The local magnetic field acting on the reference SRR is the sum of
$H_y^{M}$ and the contribution of the SRR particles: \e H_y^{\rm
loc}=H_y^{J}+C_dm=H_y^{J}+C_dabM \lab{hloc}\f 
where $C_d$ (the interaction factor of the lattice of magnetic dipoles) is given by
\cite{JOSA} \e C_d={\oo q\over 2ab\eta}+j{k^3\over 6\pi mu_0}+{\oo
\over 2ab\eta}{\sin kd\over \cos kd -\cos \beta d} \lab{bbb}\f
A similar relation has been given  in \cite{JOSA} for a lattice of
electric dipoles (the only difference as compared to \r{bbb} is the factor
$\eta^2$). Here $q$ denotes the real part of the dimensionless
interaction factor of a 2D grid of dipoles with periods $a,b$.  In
\cite{JOSA} the closed-form expression for $q$ is given for the
case $a=b$:
$$
q_0={1\over 2}\left({\cos kas\over kas}-\sin kas\right)
$$
where the number $s$ is approximately equal to $1/1.4380=0.6954$.
Relation \r{bbb} is very accurate for the case $d\gg a$, and  in the case $d=a$
its error is still quite small \cite{JOSA}.

The responses of the reference SRR and the reference wire to
the local fields can be written as
\begin{eqnarray}
m &=& a_{mm}H^{\rm loc}
\lab{refm} \\
I &=& {E^{\rm loc}\over Z_{w}}
\lab{refw}
\end{eqnarray}
where $Z_{w}$ is given by \cite{Belov} \e
Z_{w}={k\eta\over 4} \left[1-{2j\over \pi}\left( \log{kr_0\over 2}
+\gamma \right) \right] \lab{zwire}\f where $r_0$ is the effective
radius of the wire ($r_0=w/4$ if made from a strip with width $w$).

To obtain the dispersion equation we substitute formulas \r{eloc}, \r{hloc}, \r{aaa}, \r{bbb}, \r{esum}  and
\r{hsum} into \r{refm} and \r{refw}. Since $m=Mab$ and $I=Jb$
we obtain the following system of equations:
\begin{eqnarray}
M \left({1\over a_{mm}}-A-j{k^3\over 6\pi \mu_0}\right) &=&
j{BJ\over ab}
\lab{re1} \\
J (Z_w-C_w) &=& -\oo BM
\lab{re2}
\end{eqnarray}
where
$$
A={\oo \over 2ab\eta}\left({\sin kd\over \cos kd -\cos \beta d}+
q\right)
$$
and
$$
B={1\over 2}{\sin \beta d\over \cos kd -\cos \beta d}
$$
The parameter $B$ describes the interaction between the currents in
the wires and the magnetic moments of the SRR:s. $B$ is not a quasi-static
parameter even at low frequencies since it does not approach zero
at zero frequency. Its presence in the dispersion equation
strongly influences the result for the propagation constant $\beta$ at all frequencies.

%This is the answer to the first question we put in
%Introduction.

Relations \r{ima} and \r{zwire} lead to the cancellation of
the imaginary part at the left-hand side of \r{re1} and  the real part at the
left-hand side of \r{re2}. Thus, system of equations \r{re1} and \r{re2} gives
the following real-valued dispersion equation:
$$
\left({\rm Im}(Z_w)-{\rm Im}(C_w)\right)\left[{\rm Re}\left({1\over a_{mm}}\right)-A\right]=
-{\oo^2B^2\over ab}
$$
It can be re-written as the following quadratic equation with respect to
$\cos \beta d$:
$$
\left[{2ab\over \oo\eta}{\rm Re}\left({1\over
a_{mm}}\right)-q\right] (\cos kd-\cos \beta d)\sin kd
$$
\e
-\left(1+{kb\over \pi}\log{b\over 2\pi r_0}\right)\sin^2 kd
-\cos^2 \beta d+1=0 \lab{exp}\f
There are two roots $\beta_{1,2}$
for the  dispersion equation \r{exp} at each frequency. One of them is
exactly equal to $k$ (the wavenumber in the host medium). This root
corresponds to the wave with polarization $\-E=E_y\-y_0$ and
$\-H=H_x\-x_0$. This wave excites neither wires nor SRR:s and does
not interact with the structure. Another root corresponds to the
wave with polarization $\-E=E_x\-x_0$ and $\-H=H_y\-y_0$. This is
the interacting wave which is of interest. In our
dispersion curves we keep both solutions of \r{exp}.

%One of them  satisfy to the relation following from \r{re1} and
%\r{re2}
%$$
%{{\rm Im}(Z_w)-{\rm Im}(C_w)\over \oo B}={B\over ab({\rm
%Re}\left({1\over a_{mm}}\right)-A)}
%$$
%and another root does not.

\section{Dispersion curves}

As numerical examples we choose the following parameters for the
structure shown in Fig. \ref{grid}: the size of SRR particle
(outer diameter of the rings) is $D=3.8$ mm, the width of the strip
(forming the rings) is $w=1$ mm, the radius of wire cross section
is $r_0=0.2$ mm, the distance between the rings (which is chosen so that
the resonance of $a_{mm}$ is at 6 GHz) is $h=0.84$ mm.
Lattice periods $a=b=8$ mm
and $d=16$ mm are  chosen in our first example. In the second example, we choose $a=b=d=8$ mm. In the third
example, we choose $a=b=d=4$ mm.

\begin{figure}
\centering \epsfig{file=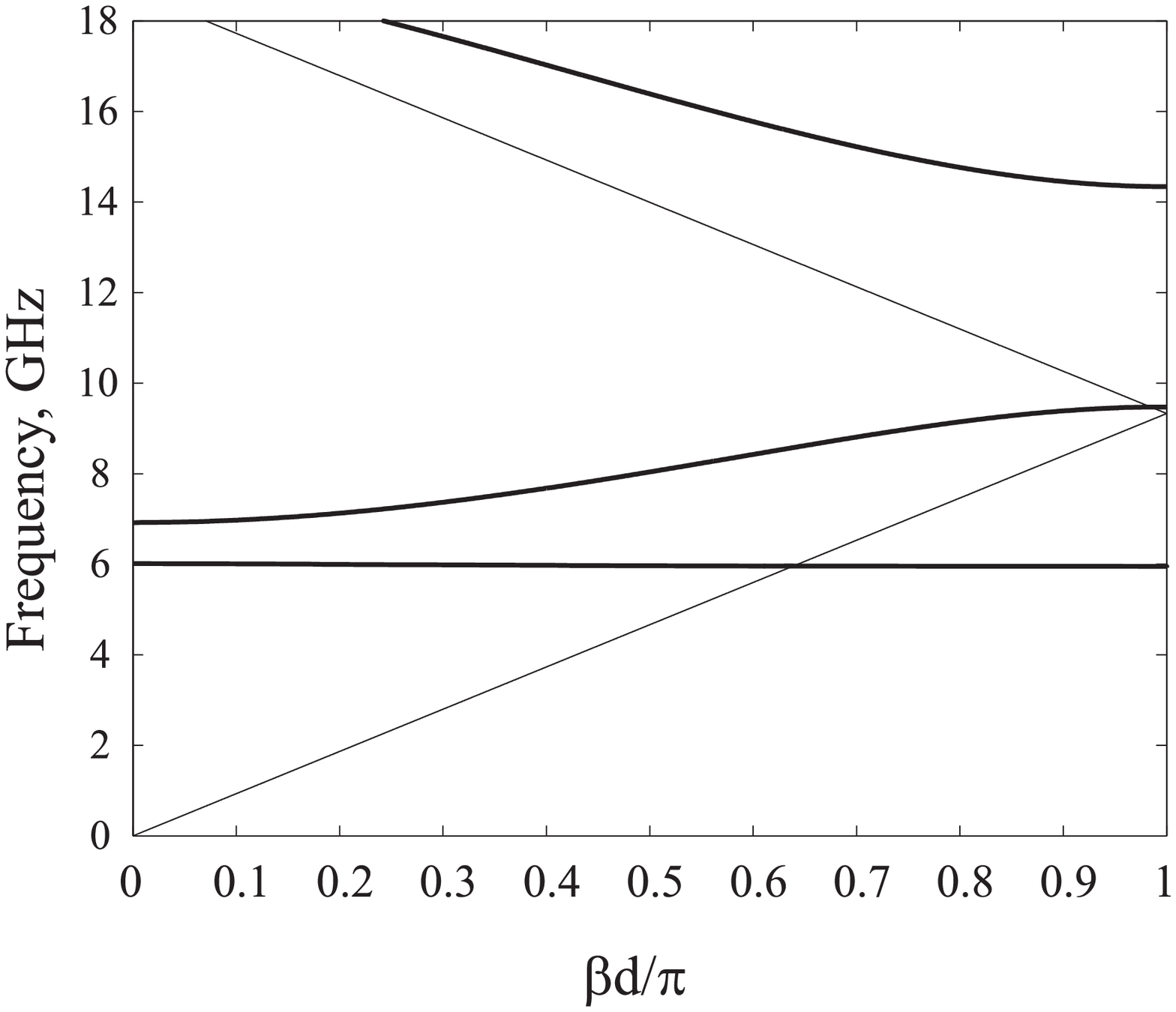, width=9cm}
\caption{Dispersion plot of the structure for $d=2a=2b=16$ mm (thick lines).
Thin lines coincide with the dispersion plot of host medium
and correspond to the wave polarization ${\bf E}=E_y{\bf
y}_0$,${\bf H}=H_x{\bf x}_0$.} \label{plot16}
\end{figure}

Fig. \ref{plot16} gives the dispersion curve in a form
commonly used in the literature of photonic crystals
(see e.g. \cite{Mead}). It represents the dependence of the eigenfrequencies on
the normalized propagation factor $\beta d/\pi$ over the first
Brillouin zone ($0<\beta <\pi/d$) for the case when $d=2a=2b=16$ mm.
Straight lines correspond to non-interacting waves. Curved
lines correspond to interacting waves. The only difference of
this curve as compared to the well-known plot for the wire medium (see e.g.
\cite{Belov} and \cite{Engheta}) is the mini-band at about 6 GHz,
in which the group velocity is negative. In this narrow frequency
band wave propagation is prohibited in the
lattice of wires. Therefore, the mini-band is due to 
the presence of SRR:s and the resonant
magnetization of the SRR lattice. The eigenfrequencies (associated with
waves with $\-E=E_x\-x_0$ and $\-H=H_y\-y_0$) within this
pass-band are shown by the crosses.

\begin{figure}
\centering \epsfig{file=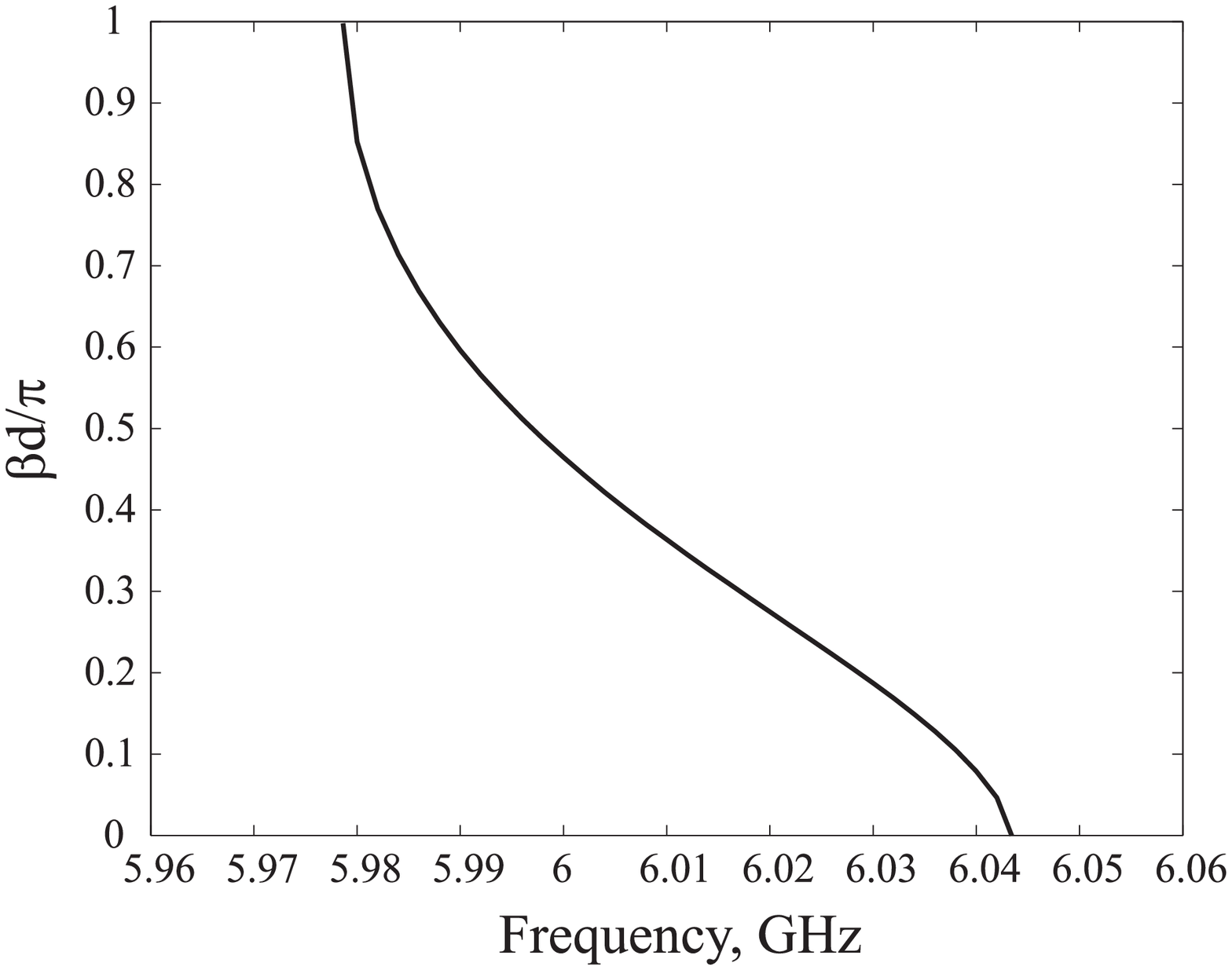, width=9cm} \caption{Normalized
propagation factor of the backward wave versus frequency
for $d=a=b=8$ mm.} \label{respas}
\end{figure}

The resonant pass-band becomes wider if the period $d$ decreases. From Fig. \ref{respas} one can see the dependence of the propagation
factor on the frequency in the vicinity of the SRR resonance for the
case $a=b=d=8$ mm. The solid line in this figure corresponds to the
non-interacting wave.

The backward-wave region corresponds to the frequencies
5.980-6.045 GHz, whereas the resonant frequency of ${\rm
Re}(a_{mm})$ given by \r{omega} is 6.000 GHz. ${\rm
Re}(a_{mm})$ becomes negative at 6.000 GHz. Thus, within the
backward-wave band ${\rm
Re}(a_{mm})$ is mainly negative. The group
velocity of the backward wave is relatively small (it
approximately equals $5.3\cdot 10^{-3}c$, where $c$ is the
speed of light).

In order to understand whether it is possible to homogenize the structure
at the frequencies when the backward-wave region exists, we
studied $\beta$ for both propagating and decaying modes.

\begin{figure}
\centering \epsfig{file=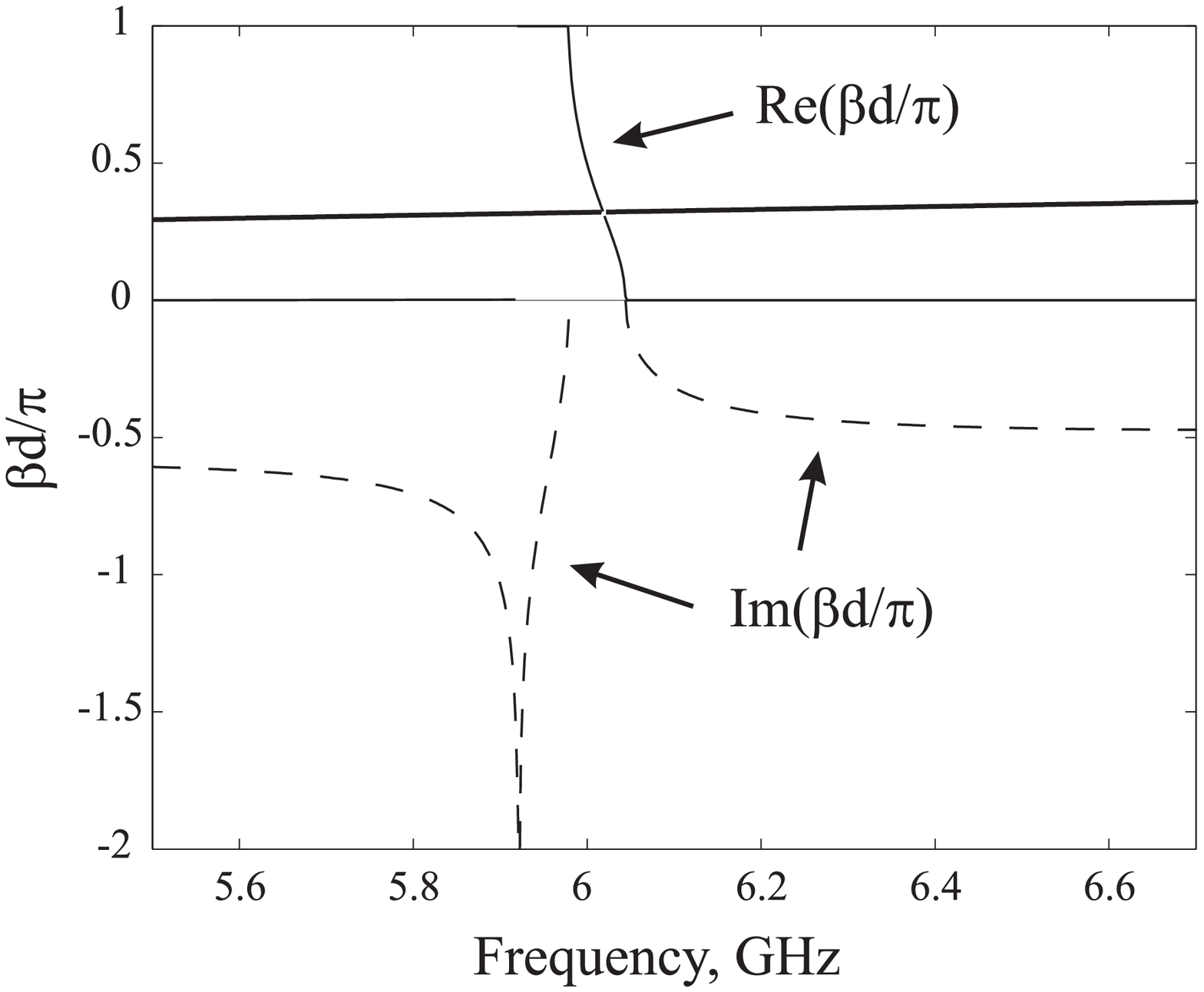, width=9cm}
\caption{Real (thin solid line)
and imaginary (dashed line) parts of normalized propagation factor
versus frequency for $d=a=b=8$ mm. Thick line corresponds to the propagation
of wave with polarization ${\bf E}=E_y{\bf y}_0$,${\bf H}=H_x{\bf x}_0$
in host material.}
\label{bothroots}
\end{figure}

Fig. \ref{bothroots} shows the frequency dependence of
both the real and imaginary parts of the normalized propagation factor
$\beta d/\pi$ for the case $a=b=d=8$ mm. It is
clear that outside the SRR resonant band  the eigenmodes of the structure  are the same as those of the wire medium \cite{Belov}. 
The thick straight line corresponds to the non-interacting wave.

From Fig. \ref{bothroots} one can see that the eigenmodes within the
frequency band 5.92-5.98 GHz are complex. The lower limit of the
backward wave region (5.980-6.045 GHz) is the upper limit of the
complex-mode band. Complex modes cannot exist in continuous media.
These modes are known for electromagnetic crystals with different geometries (see, e.g.
\cite{Belov,nonresPRE}). These are decaying modes though 
the real part of the propagation factor is ${\rm Re}\beta=\pi/d$. The existence of
this real part of $\beta$ reflects the fact that the directions of
the currents in the wires are alternating along the propagation axis
(two adjacent currents have opposite directions and this can be
interpreted as the phase shift $\pi$ between them due to the real
part of the complex propagation factor).

Therefore, the homogenization is possible within one (the upper) half of the
SRR resonant band but impossible within another (the lower) half of
the SRR resonant band (though for these frequencies the structure
periods are much smaller than the wavelength in the background
medium).

%This is the answer to the second question we put in
%Introduction.

\section{Homogenization}

Let us try to consider the structure (in the case
$a=b=d$) as a uniaxial magneto-dielectric medium. Then the interacting
wave (propagating along the $z$ axis with $\-E=E_x\-x_0$,$\-H=H_y\-y_0$)
also satisfies the following constitutive equations:
$$
D_x=\epsilon_0E_x+P_x^{\rm bulk}=\epsilon_0\epsilon_{xx}E_x
$$
$$
B_y=\mu_0H_y+M_y^{\rm bulk}=\mu_0\mu_{t}H_y
$$
Define the following ratio: \e \alpha=\eta{P_x^{\rm bulk}\over M_y^{\rm
bulk}}={1\over\eta} {(\epsilon_{xx}-1)\over {(\mu_t-1)}}{E_x\over
H_y} \lab{rat}\f where $E_x$ and $H_y$ are the field components 
averaged over the cubic cell $a\x a\x a$. $P_x^{\rm bulk}$ and
$M_y^{\rm bulk}$ are the bulk electric and magnetic polarizations
related with the surface current $J$ and surface magnetic
polarization:
$$
P_x^{\rm bulk}={J\over j\oo d}\qquad M_y^{\rm bulk}={M\over d}
$$
From Maxwell's equations we easily obtain \e {E_x\over H_y}=\eta
\sqrt{\mu_t\over \epsilon_{xx}} \lab{imped}\f

Substituting  \r{aaa} and \r{zwire} into \r{re2}, we
obtain: 
$$ \alpha(\oo)= {\eta J\over j\oo M}= {\eta B\over
j(Z_w-C_w)}=
$$
\e
-\displaystyle \frac{\pi\sin \beta d}{\pi\sin
kd + kb\log{\displaystyle \frac{b}{2\pi r_0}(\cos kd-\cos \beta d)}}
 \lab{alp}\f
From \r{rat} and \r{imped} it follows that \e \alpha=
{(\epsilon_{xx}-1)\over {(\mu_t-1)}}\sqrt{\mu_t\over
\epsilon_{xx}} \lab{alpha}\f
In the above equation $\beta(\oo)$ is already known from the dispersion curve and 
$k=\oo\sqrt{\epsilon_0\mu_0}=\oo/c$.

Equating the propagation factor $\beta$ to the value $\oo
\sqrt{\epsilon_0\mu_0\epsilon_{xx}\mu_t}$, we obtain \e
\mu_t={\beta^2\over k^2\epsilon_{xx}} \lab{mu}\f Substituting this
expression for $\mu_t$ into \r{alpha}, we obtain: \e
\epsilon_{xx}(\oo)={{c^2\beta^2(\oo)\over \oo^2}+ {c\beta\over \oo\alpha(\oo)}\over 1+
{\beta(\oo)\over k\alpha(\oo)}} \lab{eps}\f After 
$\epsilon_{xx}$ is found, we then evaluate $\mu_t$ through (cf. \r{eps} and \r{mu}) \e\mu_t(\oo)=
{1+{c\beta(\oo)\over \oo\alpha(\oo)}\over 1+{\oo\over c\beta(\oo)\alpha(\oo)}} \lab{mumu}\f

We have taken into account the non-local interaction in the
structure in formulas \r{eps} and  \r{mumu} though the effective permittivity and permeability are introduced
as the parameters relating $\-B,\-D$ with $\-E,\-H$ at the same
point. Therefore, unlike formulas  \r{Pend1} and \r{Pend2}, our material
parameters are not quasi-static.

\begin{figure}
\centering \epsfig{file=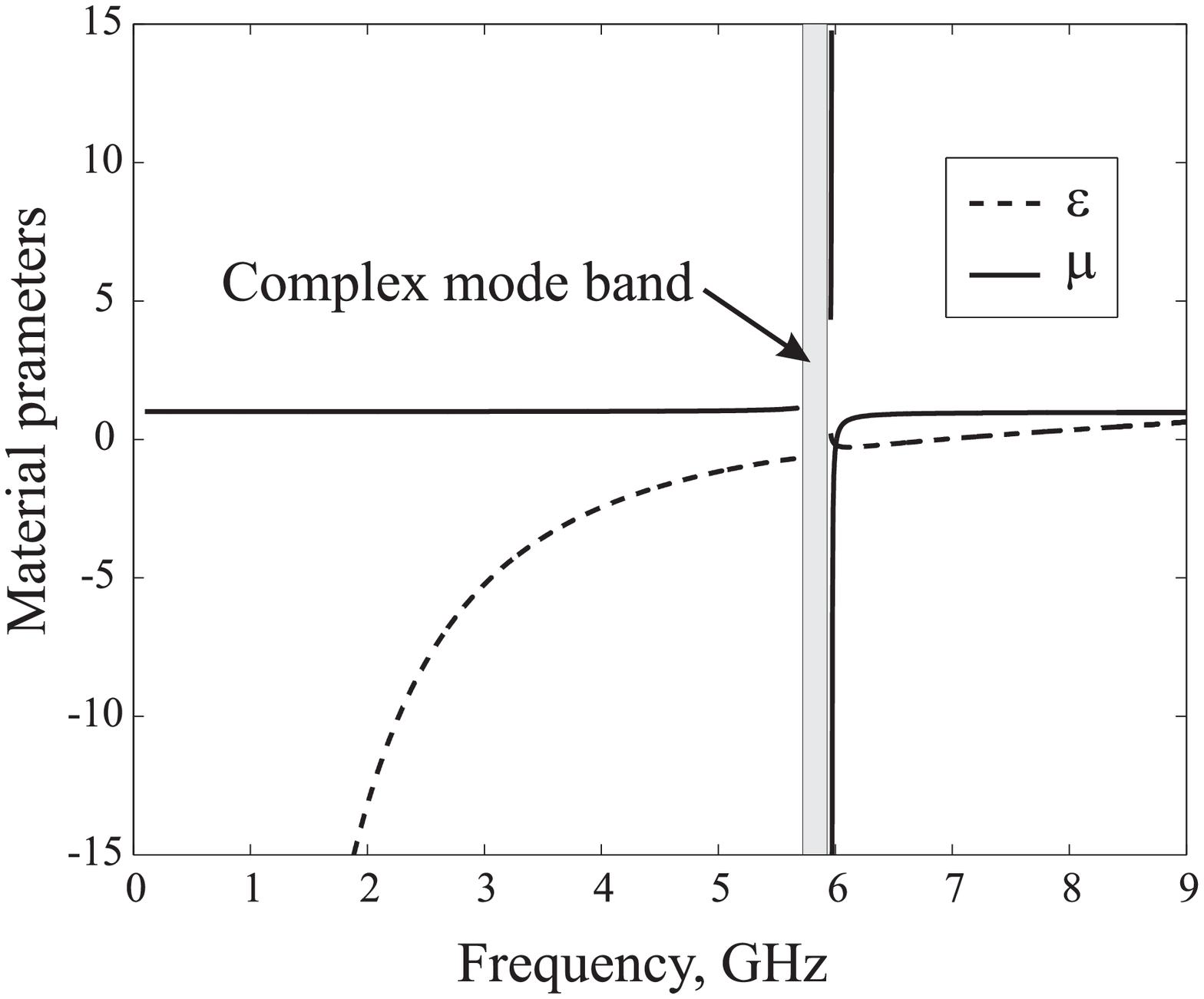, width=9cm} \caption{The axial
permittivity (dashed line) and the transversal permeability (solid line)
for the case $a=b=d=8$ mm.} \label{wide}
\end{figure}

Frequency dependencies of both $\epsilon_{xx}$
and $\mu_t$ are shown in Fig. \ref{wide} for the case $a=b=d=8$ mm. Outside the
resonant band of the SRR particles the frequency dependence of the
permittivity repeats the known result for wire media (treated as
artificial dielectric media) \cite{Brown}. The permeability is practically
equal to the unity outside the resonant band of SRR:s. Within the
complex-mode band the homogenization is forbidden \cite{Sipe,nonresPRE} and this frequency region 
is removed in this figure (both $\epsilon_{xx}$ and
$\mu_t$ calculated through \r{eps} and \r{mu} are complex within
this band).

\begin{figure}
\centering \epsfig{file=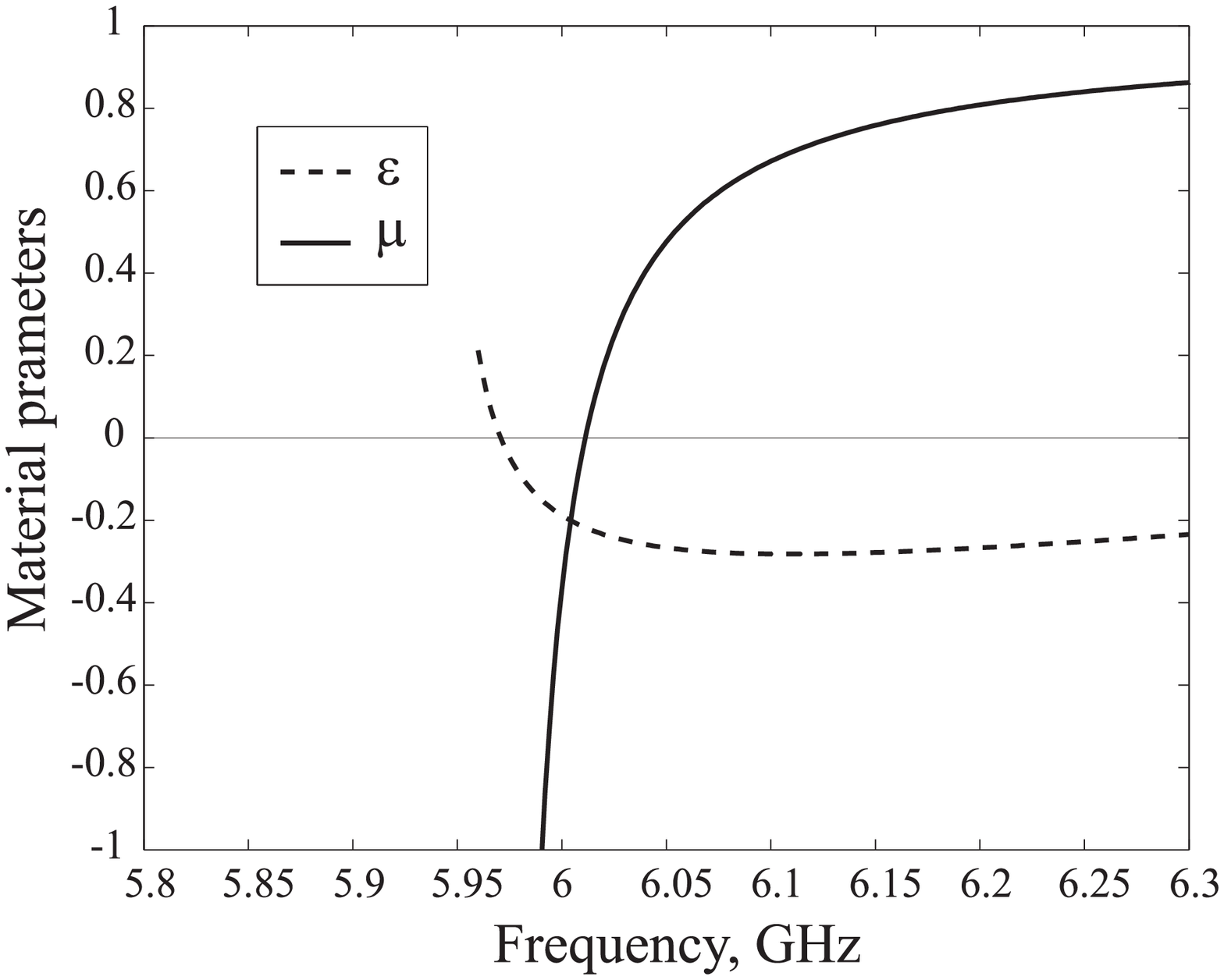, width=9cm} \caption{The axial
permittivity (dashed line) and the transversal permeability (solid line)
for the case $a=b=d=8$ mm within the SRR resonant band.}
\label{wide2}
\end{figure}

Let us consider the resonant frequency behavior of the material parameters in
details. Fig. \ref{wide2} shows the same curves as those in Fig. \ref{wide},
however, in another scale starting from the upper limit of the
complex-mode band. From Fig. \ref{wide2} one sees that the permittivity and
permeability are both negative between 5.970 and 6.020 GHz. From
Fig. \ref{respas} it follows that the backward wave propagates
between 5.980 GHz and 6.045 GHz. Thus, the backward-wave region
almost coincides with the region where both the permittivity and
permeability  are
negative. Note that this coincidence is only approximate.

Also we have indicated in Fig. \ref{wide2} the point at which permittivity and permeability 
are equal (at about 6.005 GHz). At this frequency, the medium is impedance-matched with the free space (this is useful for some applications) and the values of 
$\epsilon_{xx}$ and $\mu_{t}$ are not very high (the homogenization is then allowed).

As a main result, one can see from Fig. \ref{wide2}  that the
permittivity does not follow (within the
resonant band)  the law (even qualitatively) suggested in \cite{negative},
\cite{Science} and \cite{Smith}. it's the frequency dependence of the permittivity 
is non-monotonous (from Fig. \ref{wide2} one sees that $\epsilon_{xx}$ decreases over 5.96-6.09 
and increases after 6.090 GHz as the frequency increases).

In the theory of continuous media one can prove that both the permittivity and
permeability  must grow as the frequency increases in the lossless
case
\cite{Landau}. In our case the permeability grows everywhere as the frequency increases 
(until the first spatial resonance of the lattice, i.e.,  $kd=\pi$ when it
loses the physical meaning). Thus, the frequency behavior of
the permeability is normal. However, the permittivity grows as the frequency increases 
only at
the frequencies when the magnetization of SRR is small and the
interaction of the SRR:s and wires is negligible. Within the band of the
backward wave the permittivity  decreases as the frequency increases. Therefore, the
homogenization procedure we have developed is not completely
consistent with the theory of \cite{Landau}. The reason for this
disagreement is that the material parameters considered in  \cite{Landau}
are quasi-static (i.e., the polarization of the medium at
a given point is determined by the field at this point) while our model 
takes into account the non-local interaction of
the SRR lattice and the  wire lattice. We found that the visible
difference between the quasi-static model and our model
is within the SRR resonant band. However, the influence of
the non-local interaction is revealed in the permittivity of the wire
lattice disturbed by the presence of the SRR:s.

The lattice of infinite wires is spatially dispersive at
all frequencies since the wires are longer than any possible
wavelength. When the wave propagates strictly in the plane
orthogonal to the axis of the wires one can still neglect the spatial
dispersion since all parameters are independent of  the $x$-coordinate.   Thus, the
problem is   two-dimensional  and possible to be homogenized \cite{Brown}. However, if there is
a lattice of scatterers with which the wires interact, the
situation becomes quite different (even for propagation orthogonal
to the wires). Here the problem is not two-dimensional  and the
wire current is influenced by all the SRR particles positioned
along its infinite length. It results in the abnormal frequency
behavior of the effective permittivity of the  structure.

\section{Conclusion}

In the present paper we have developed an analytical model for a structure similar to the one for which the negative
refraction at microwave frequencies was first observed (formed by combined lattices of infinitely long wires and split-ring resonators) \cite{Science}.  We have derived a self-consistent
dispersion equation and studied the dispersion properties of the
lattice.  The explicit dispersion
equation clearly confirms the existence of the narrow
pass-band within the resonant band of the split-ring resonators. In this pass-band the group and phase velocities of the
propagating wave are in opposite directions (i.e., a backward wave). Negative refraction is the direct
consequence of the backward wave  (instead of the negative values of permittivity and permeability as thought in many works \cite{Garcia}). Thus, the experimental observations of
\cite{Science}  are consistent to  our model.

Our model also allows the homogenization of the composite structure.
The obtained dispersion curves have been used to
calculate correctly the effective permittivity and permeability in
the frequency band where the structure can be homogenized. It is
interesting to see that the dispersion curves agree well with the
Veselago theory which predicts  backward waves when both
permittivity and permeability are negative. Outside the resonant band of the SRR particles, the effective  permittivity of the whole structure is the same as that of the wire
lattice and the effective permeability
is equal to 1. However, within the SRR resonant band,
there is a sub-band where the homogenization is forbidden since
the complex mode  satisfies the dispersion equation at
these frequencies. We found that the frequency region in which  both $\epsilon$ and
$\mu$ are negative coincides approximately with the backward
wave band. In this region the frequency dependence of
the effective permittivity is abnormal. We interpret this as the result of the
low-frequency spatial dispersion which is inherent for the wire
medium in the presence of the resonant scatterers.

\section*{Acknowledgement}

The support of C.R. Simovski by Swedish Royal Academy of Science
(KVA) and Russian Foundation for Basic Research (grant 01-02-16856) is
gratefully acknowledged. P.A. Belov thanks SUMMA Foundation for support of his studies by SUMMA Graduate Fellowship in Advanced Electromagnetics.

\end{document}